# Comparing entropy with tests for randomness as a measure of complexity in time series

**Chee Chun Gan**  cg8pa@viriginia.edu
*Department of Systems and Industrial Engineering*
*University of Virginia*

**Gerard Learmonth**  jl5c@virginia.edu
*Center for Leadership Simulation and Gaming*
*Center for Large-Scale Computational Modelling*
*Frank Batten School of Leadership and Public Policy*
*University of Virginia*

**Abstract**

Entropy measures have become increasingly popular as an evaluation metric for complexity in the analysis of time series data, especially in physiology and medicine. Entropy measures the rate of information gain, or degree of regularity in a time series e.g. heartbeat. Ideally, entropy should be able to quantify the complexity of any underlying structure in the series, as well as determine if the variation arises from a random process. Unfortunately current entropy measures mostly are unable to perform the latter differentiation. Thus, a high entropy score indicates a random or chaotic series, whereas a low score indicates a high degree of regularity.

This leads to the observation that current entropy measures are equivalent to evaluating "how random" a series is, or conversely the degree of regularity in a time series. This raises the possibility that existing tests for randomness, such as the runs test or permutation test, may have similar utility in diagnosing certain conditions.

This paper compares various tests for randomness with existing entropy-based measurements such as sample entropy, permutation entropy and multi-scale entropy. Our experimental results indicate that the test statistics of the runs test and permutation test are often highly correlated with entropy scores and may be able to provide further information regarding the complexity of time series.

## 1.    Introduction

Entropy measures have gained widespread use in the analysis of complex real-world data. The term "entropy" first originated in the field of thermodynamics and can be interpreted as the amount of information needed to completely specify the physical state of a system. A very orderly and regular system has a low entropy value. An example of this is a system consisting of a container of hydrogen and helium molecules where all the hydrogen molecules are on one side of a divider and all the helium molecules are on the other side. In contrast, a system where the hydrogen and helium molecules are uniformly distributed throughout the container has very high entropy as the position of each molecule has to be completely specified in order to describe the state of the system.

The concept of entropy was further developed in the field of non-linear dynamic analysis and chaos as a measure of the complexity of a system. In Shannon's [1] seminal work on information theory, he defined entropy as the "information content" of a system. However, the concept of entropy remained largely theoretical until Pincus [2] developed Approximate Entropy (ApEn) as a measure of changing complexity which could be applied to real-world data sets. Following on from Pincus's work, various other entropy measures have been proposed for the same purpose. Richman and Moorman [3] introduced Sample Entropy (SampEn), a modified version of ApEn, to correct for the self-match bias in ApEn and to improve on several other statistical properties. Bandt and Pompe [4] proposed Permutation Entropy (PermEn) as an alternative measure of complexity for time series. Costa et. al. [5] developed Multi-Scale Entropy (MSE) to account for structural interactions across multiple time scales.

However, the abovementioned entropy measures all share a common attribute in that a maximal entropy score is assigned to completely random data, i.e. white noise. In that sense, entropy can be considered to be a measure of the degree of regularity in data where the presence of underlying structure will reduce the entropy score from the maximal value.

By extending on this premise, it should be possible to obtain similar information by utilizing existing statistical tests for randomness such as the Runs test and permutation test [6]. The rest of this paper explores the efficacy of such tests as compared to the abovementioned entropy measures. Section 2 outlines the basic framework of each entropy measure and test for randomness that was utilized in the experimental runs. Section 3 contains the experimental results for various test data sets, while section 4 discusses the conclustions from the experiments.

## 2. Experimental methods

### 2.1 Entropy measures

Sample Entropy(m,r)

As presented by Lake, Richman, Griffin and Moorman in their study of neonatal heart rates [7], SampEn is the negative natural logarithm of the conditional probability that a dataset of length m, given that it has repeated itself for m points (within a tolerance limit r that is commonly based on the standard deviation of the data), will repeat itself for m+1 points. SampEn can be calculated using the following equation:

$$SampEn = -log\frac{A}{B}$$

where A is the number of pairs of vector subsets of length m+1 which have a distance function less than r, while B is the number of pairs of vector subsets of length m which similarly have a distance function less than r. The main difference between SampEn and ApEn is that SampEn does not allow self-matching of points while ApEn does, meaning that ApEn always has a value of at least 1 for A and B.

For our experiments, the parameter m = 2 was chosen and r was set to be 0.2 times the standard deviation of the test data. A SampEn score of 0 indicates linear or highly regular data, while randomly generated data returns a SampEn score between 2.2 and 2.3.

Permutation Entropy(n)

PermEn was developed to handle the presence of noise in real-world data. For a time series $\{x_0, ..., x_{N-1}\}$ the PermEn algorithm splits the data into overlapping n-tuples, where n is the embedding dimension. Each n-tuple is then sorted in ascending order, which generates a "permutation type" π according to the ordering of the sorted data. As an example, consider the 3-tuple $\{x_0, x_1, x_2\}$ = {3,5,1}. The sorted tuple is $\{x_2, x_1, x_0\}$ which leads to a π = 2,1,0. For embedding dimension n there are n! possible permutation types. The relative frequency $p(\pi_i)$ is determined for each $\pi_i$, for $1 \leq i \leq n!$, according to the following equation:

$$p(\pi_i) = \frac{number\ of\ occurrences\ of\ type\ \pi_i}{N - n + 1}$$

The permutation entropy H(n) is then calculated as follows:

$$H(n) = -\sum_{i}^{n!} p(\pi_i) \log p(\pi_i)$$

H(n) ranges from 0 to log n!, with 0 indicating a series that is monotonically increasing or decreasing and log n! indicating a completely random series. In the experimental portion, H(n) is rescaled by dividing by log n!, thus normalizing H(n) to return values between 0 and 1 with 0 indicating highly regular data and 1 indicating maximal entropy. The parameter n = 5 was used for the calculation of H(n).

Multi-Scale Entropy (m,n)

In their paper [5], Costa, Goldberger and Peng observed that most entropy measures only consider a one-step difference and thus only measure entropy based on the smallest scale. Multi-Scale Entropy (MSE) down-samples the original time series {$x_1$, …, $x_N$} according to a scale factor T. The new time series {$y^T$} is obtained using the formula $y_j^T = \frac{1}{T} \sum_{i=(j-1)T+1}^{jT} x_i, 1 \leq j \leq \frac{N}{T}$. In effect, the original series is partitioned into N/T disjoint sets. The mean of each disjoint set then forms a data point in the new series {$y^T$}. Sample entropy is then calculated while varying T and the resultant values plotted against T.

Since the MSE methodology revolves around the resampling process, in our experiments the resampling process was used before applying the various methodologies examined. Scale factors {2,3,4,5,10} were evaluated, with scale factor 1 being excluded as the down-sampled series would be identical to the original series.

## 2.2 Tests for randomness

Permutation test (t)

The permutation test for randomness [6] should not be confused with the permutation tests involving reshuffling of data to obtain more accurate test statistics. The permutation test for randomness is performed by first partitioning the original time series into groups of t elements. In the event that the original time series is not perfectly divisible by t, the remaining data points are discarded. The elements of each group are then sorted to obtain an ordering of the element indices. As there are $t!$ possible orderings in each group, a chi-square test can then be performed with t! categories and the assumption that the probability of each distinct ordering is $1/{t!}$. The chi-square statistic is interpreted as the distance from the expected value given the null hypothesis that the input data is uniformly distributed.

Thus, a high value indicates a high degree of regularity (conversely, departure from randomness) while a low value indicates a high likelihood of the null hypothesis being true.

It should be noted that the algorithm for the permutation test for randomness is very similar to that for calculating permutation entropy as presented in section 2.2. The main differences are that the partitions in the permutation test do not overlap, and that instead of summing the chi-square test statistic for each permutation the permutation entropy algorithm calculates H(n). In our experiments, a partition size of t=5 was chosen, in part to be consistent with the tuple size for permutation entropy.

Runs test

The runs test [6] examines the time series for the length of sequences that increase or decrease monotonically. The underlying basis for the runs test is that a non-random series will tend to have either more or less frequent runs than expected under a purely random distribution. For the purposes of our experimental analysis, the R function runs.test() in the "lawstats" package [8] was used to calculate the two-sided runs test statistic and p-values.

Similar to the permutation test statistic, the runs test statistic can also be interpreted as the distance from the expected value given a null hypothesis of a random originating distribution. The greater the difference from zero, the greater the degree of regularity in the series. Conversely, a value close to zero indicates a high probability of the null hypothesis being true.

### 3. Experimental results

Entropy measures are primarily used to provide quantitative comparisons between various time series. By themselves, the various entropy scores are not sufficient to determine if a series is chaotic. However, entropy scores can be used to rank multiple time series according to the degree of regularity exhibited.

Conversely, tests for randomness have traditionally been used to provide probabilistic likelihoods of whether the input series is randomly distributed. Thus, the main focus of these tests has been on the p-values, or probability of encountering such an input series given the assumption that the null hypothesis is true (the series is randomly distributed) rather than the actual values of the test statistic. However, when used as a comparative measure among multiple time series the test statistic may also be able to give pertinent information, in a fashion similar to comparing entropy scores.

In our experiments, we evaluated the performance of Sample Entropy (SampEn), Permutation Entropy (PermEn), the permutation test chi-square test statistic (p.test) and the runs test statistic (runs.test). In certain cases, the Multi-Scale Entropy (MSE) framework was applied by downsampling the original series and evaluating the changes in the various metrics as the scale factor T increases.

## 3.1 Random data

The first test performed compared three time series (each with 1000 data points) generated from random distributions. The first series was generated from a Uniform(0,1) distribution, the second series from a Normal(0,1) distribution, and the third series from an Exponential(1) distribution. In each case, 30 replications of 1000 data points were generated and the mean score recorded. Figures 1a to 1c below show the resulting scores from SampEn, PermEn, the permutation test chi-square statistic and the runs test statistic. All scores for the various metrics are rescaled to be between 0 and 1 for better comparison. In addition, the inverse of the natural log is applied to the chi-square statistic from the permutation test, while the inverse of the absolute runs test statistic is shown. This was done to invert the plots to better correspond with the interpretation of the entropy measures, i.e. a high value corresponds to low regularity and a low value corresponds to high regularity. Detailed results of the tests, including p-values, can be found in Table 1.

Figure 1a                                                     Figure 1b

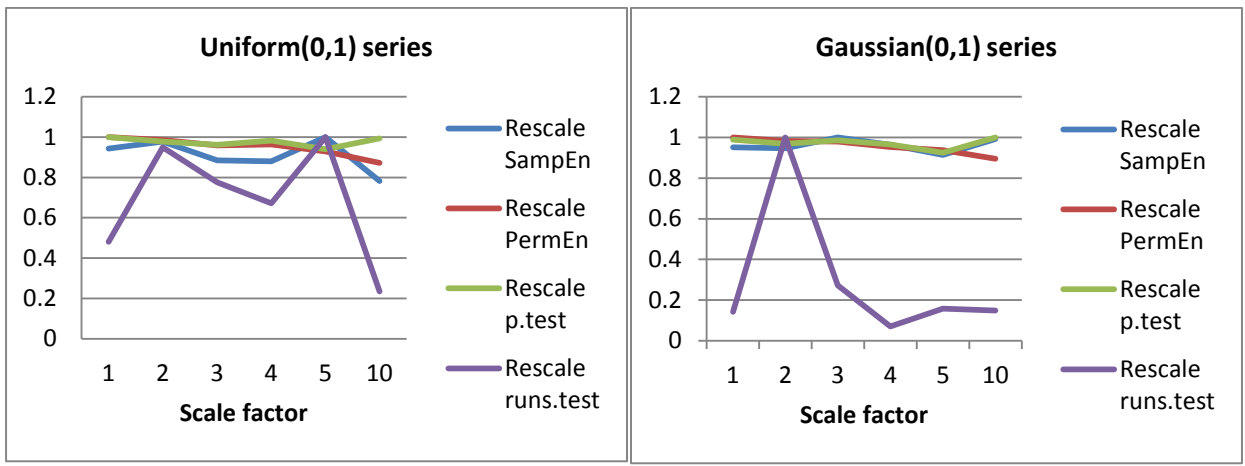

Figure 1c

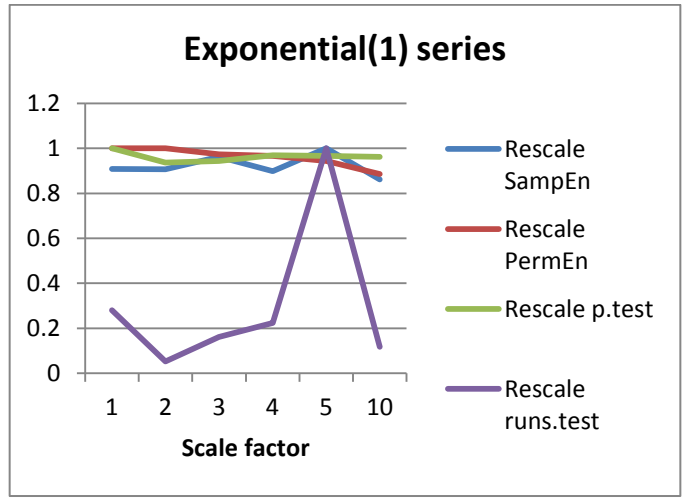

Table 1 : MSE analysis for random data

| Uniform | | | | | | |
|---|---|---|---|---|---|---|
| Scale factor | 1 | 2 | 3 | 4 | 5 | 10 |
| **SampEn(2)** | 2.238808 | 2.32396 | 2.100495 | 2.089392 | 2.374906 | 1.856298 |
| **PE(5)** | 0.987112 | 0.973145 | 0.946222 | 0.951224 | 0.917653 | 0.860831 |
| **Perm test(5)** | 108.3935 | 120.7855 | 130.3399 | 117.9717 | 145.9562 | 111.9328 |
| **Perm test(5) p-val** | 0.7471 | 0.437094 | 0.224877 | 0.509409 | ***0.04717*** | 0.664233 |
| **Runs test** | -0.8859 | -0.44766 | 0.547999 | -0.63373 | -0.42533 | -1.80916 |
| **Runs test p-val** | 0.3757 | 0.654397 | 0.583693 | 0.526258 | 0.670593 | 0.070426 |
| **Normal** | | | | | | |
| Scale factor | 1 | 2 | 3 | 4 | 5 | 10 |
| **SampEn(2)** | 2.168564 | 2.155924 | 2.27835 | 2.197225 | 2.083466 | 2.261763 |
| **PE(5)** | 0.988476 | 0.971931 | 0.968727 | 0.942845 | 0.92731 | 0.884947 |
| **Perm test(5)** | 117.9929 | 130.3844 | 119.4328 | 132.3682 | 163.9508 | 111.9328 |
| **Perm test(5) p-val** | 0.5089 | 0.224066 | 0.471603 | 0.189728 | ***0.003987*** | 0.664233 |
| **Runs test** | -0.6328 | 0.089532 | -0.3288 | -1.26746 | -0.56711 | -0.60305 |
| **Runs test p-val** | 0.5269 | 0.928659 | 0.742307 | 0.204991 | 0.570638 | 0.546473 |
| **Exponential** | | | | | | |
| Scale factor | 1 | 2 | 3 | 4 | 5 | 10 |

| SampEn(2) | 1.669779 | 1.665106 | 1.768694 | 1.650423 | 1.836711 | 1.581786 |
|---|---|---|---|---|---|---|
| PE(5) | 0.98541 | 0.985588 | 0.959452 | 0.953136 | 0.930561 | 0.873627 |
| Perm test(5) | 93.9944 | 127.9846 | 123.0685 | 108.374 | 109.967 | 111.9328 |
| Perm test(5) p-val | 0.9561 | 0.270497 | 0.380607 | 0.74757 | 0.711414 | 0.664233 |
| Runs test | -0.5062 | -2.68597 | -0.8768 | 0.633729 | -0.14178 | -1.20611 |
| Runs test p-val | 0.6127 | ***0.007232*** | 0.380596 | 0.526258 | 0.887255 | 0.227776 |

It can be seen that both PermEn and the permutation test show very high degrees of randomness for all three series. For the exponential series, while the rescaled SampEn score is consistently high through all scale factors, the absolute score shows a decrease in complexity when compared to the Uniform and Gaussian series. In most cases the p-values for both the permutation test and the runs test are well above common significance thresholds, correctly indicating that these series cannot be rejected as being random. However, there are several outliers highlighted in Table 1 where the p-values are below 0.05, which would lead to a rejection of the null hypothesis at a 0.05 significance level. In particular, there is significant variation in the runs test statistic compared to the other three test metrics.

We were unable to replicate the results of Costa et. al. [5], which found that the SampEn score monotonically decreases as T increases for series containing pure white noise. For SampEn, the permutation test statistic and the runs test statistic there was no discernible trend for varying T. However, the PermEn score did decrease monotonically as T increased.

## 3.2 Logistic Map

The second set of time series used was generated using the logistic map $x_{n+1} = rx_n(1-x_n)$. The orbit of the logistic map is shown below in Figure 2 for values of r ranging from 1.5 to 4. Five separate time series were generated from the logistic map. The first three series consist of 1000 points generated using x=0.3, with r taking values 3.5, 3.7 and 3.9 respectively. To remove transient effects, 5000 points were generated and the last 1000 used for the analysis.

Figure 2 : Logistic map for increasing r

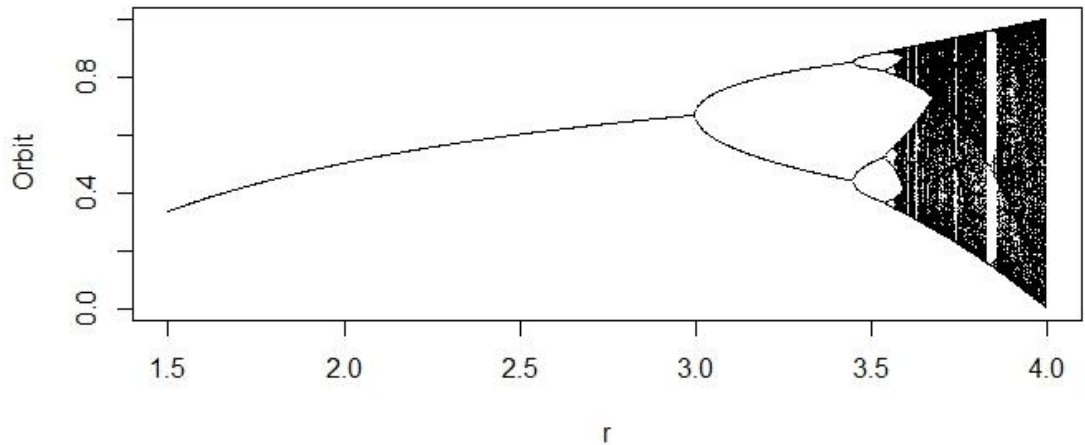

As can be seen from Figure 2, with r=3.5 the series has period 4, while for r=3.7 and r=3.9 the series results in deterministic chaos and appears much more random. Theoretically, the series with r=3.9 should exhibit a higher degree of chaos compared to the series with r=3.7. For the last time series in the set, the data from the periodic series with r=3.5 was used as the base and a random Gaussian(0,0.1) noise was added to each of the points. The plots of these four time series are shown in Figure 3.

Figure 3 : Time series from logistic map

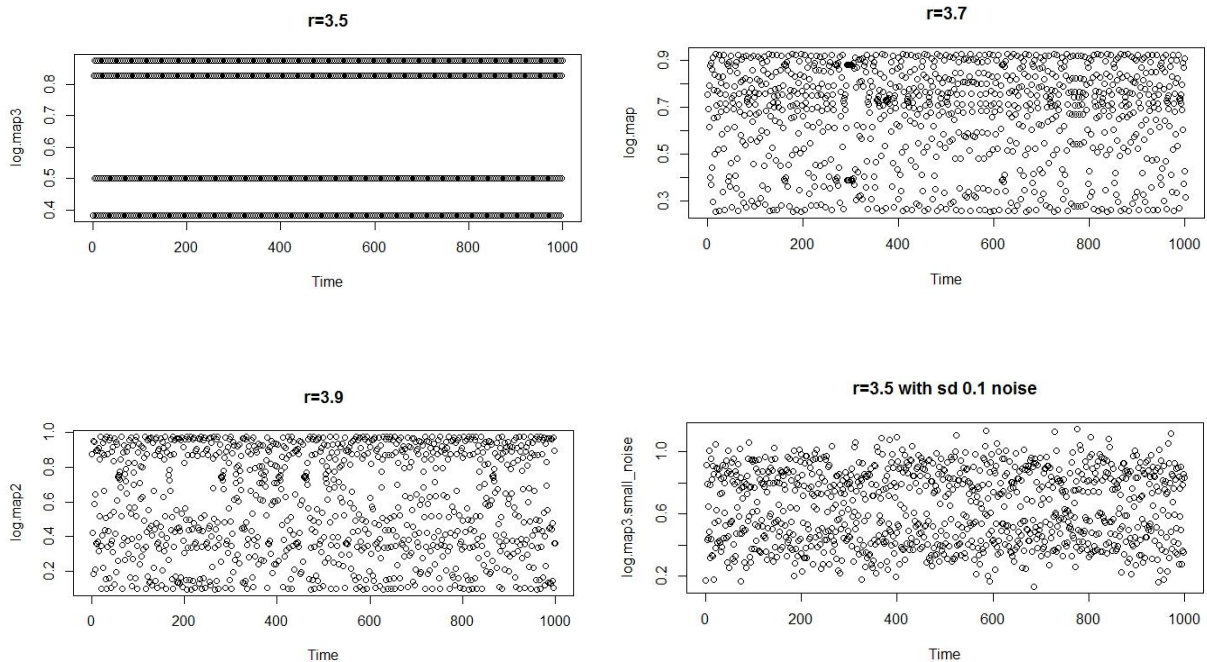

The same scores from the preceding section were calculated for these 4 time series, with the same parameter values. Figure 4 shows the rescaled scores while the raw data is provided in Table 2.

Figure 4 : Comparison for logistic map

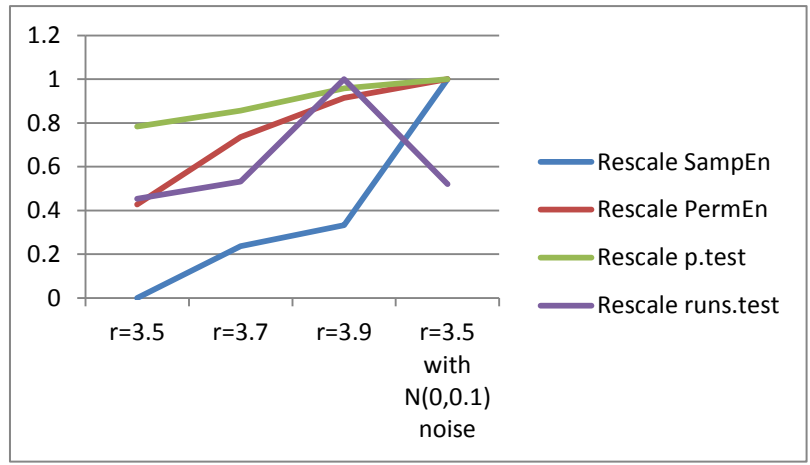

Table 2 : Scores for logistic map

|  | r=3.5 | r=3.7 | r=3.9 | r=3.5 with N(0,0.1) noise |
|---|---|---|---|---|
| **SampEn(m=2)** | 0.0000 | 0.3479 | 0.4883 | 1.4431 |
| **PermEn(5)** | 0.2896 | 0.4978 | 0.6185 | 0.6781 |
| **Perm. test(5)** | 5799.6520 | 2781.8331 | 1200.3280 | 936.3438 |
| **Perm. test(5) p-val** | 0.0000 | 0.0000 | 0.0000 | 0.0000 |
| **Runs test** | 31.5753 | 26.9561 | 14.3252 | 28.2849 |
| **Runs test p-val** | 0.0000 | 0.0000 | 0.0000 | 0.0000 |

SampEn, PermEn and the permutation test chi-square statistic are all able to detect the increase in chaos in the series. The runs test statistic does not display as much consistency as the other 3 measures, but the p-values for both the permutation test and the runs test enable us to easily reject the null hypothesis of a random process for all four series.

The MSE framework was then applied to the data sets with r=3.7 and r=3.5 with Gaussian(0,0.1) noise to evaluate the effects of increasing scale factor. Figure 5 shows that all four metrics generally vary in the same fashion for increasing scale factor.

Figure 5

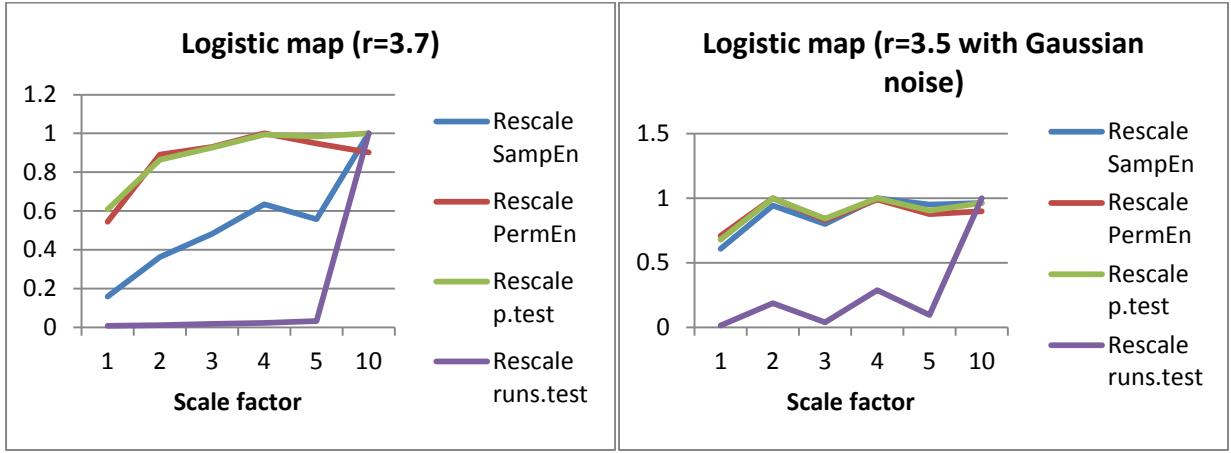

Table 3: MSE analysis for logistic map

| Logistic map r=3.7 | | | | | | |
|---|---|---|---|---|---|---|
| **Scale factor** | 1 | 2 | 3 | 4 | 5 | 10 |
| SampEn(2) | 0.3479 | 0.7899 | 1.0515 | 1.3852 | 1.2181 | 2.1832 |
| PE(5) | 0.4978 | 0.8134 | 0.8494 | 0.9134 | 0.8667 | 0.8239 |
| Perm test(5) | 2781.8331 | 262.3685 | 181.2398 | 127.5694 | 133.9598 | 123.9256 |
| Perm test(5) p-val | 0.0000 | 0.0000 | 0.0002 | 0.2791 | 0.1649 | 0.3601 |
| Runs test | 26.9561 | -17.2798 | 11.2888 | -9.1257 | 6.2382 | 0.2010 |
| Runs test p-val | 0.0000 | 0.0000 | 0.0000 | 0.0000 | 0.0000 | 0.8407 |
| Logistic map r=3.5 with N(0,0.1) noise | | | | | | |
| **Scale factor** | 1 | 2 | 3 | 4 | 5 | 10 |
| SampEn(2) | 1.4431 | 2.2351 | 1.8983 | 2.3735 | 2.2532 | 2.2824 |
| PE(5) | 0.6781 | 0.9561 | 0.7941 | 0.9449 | 0.8389 | 0.8578 |
| Perm test(5) | 936.3438 | 103.9875 | 246.6824 | 103.5751 | 169.9490 | 123.9256 |
| Perm test(5) p-val | 0.0000 | 0.8349 | 0.0000 | 0.8421 | 0.0015 | 0.3601 |
| Runs test | 28.2849 | 2.1488 | 10.3024 | -1.3942 | 4.2533 | -0.4020 |
| Runs test p-val | 0.0000 | 0.0317 | 0.0000 | 0.1633 | 0.0000 | 0.6877 |

The MSE analysis in table 3 shows a general increase in each metric as the scale factor increases. This is to be expected as the logistic map is originally deterministic. The downsampling procedure breaks the correlation between successive points, leading to a loss of regularity. As the logistic map with r=3.5 is periodic with period 4, the results for even scale factors (2,4) can be seen to differ greatly from the results for odd scale factors (1,3,5). If the results are separated into even and odd categories, it is again evident that increasing the scale factor leads to a decrease in the regularity of the series.

### 3.3    Santa Fe Time Series Competition – Set A

A univariate time series was obtained from set A of the Santa Fe Time series competition [9]. The time series consists of 1000 intensity measurements from a laser in a physics experiment that varies from periodic to chaotic pulsations. Additional time series were generated by adding a Gaussian noise component with mean 0 and standard deviation equal to the standard deviation of the original laser intensity series multiplied by 0.1, 0.2 and 1. Plots of the four time series are shown in Figure 6 below.

Figure 6 : Santa Fe time series competition – Set A

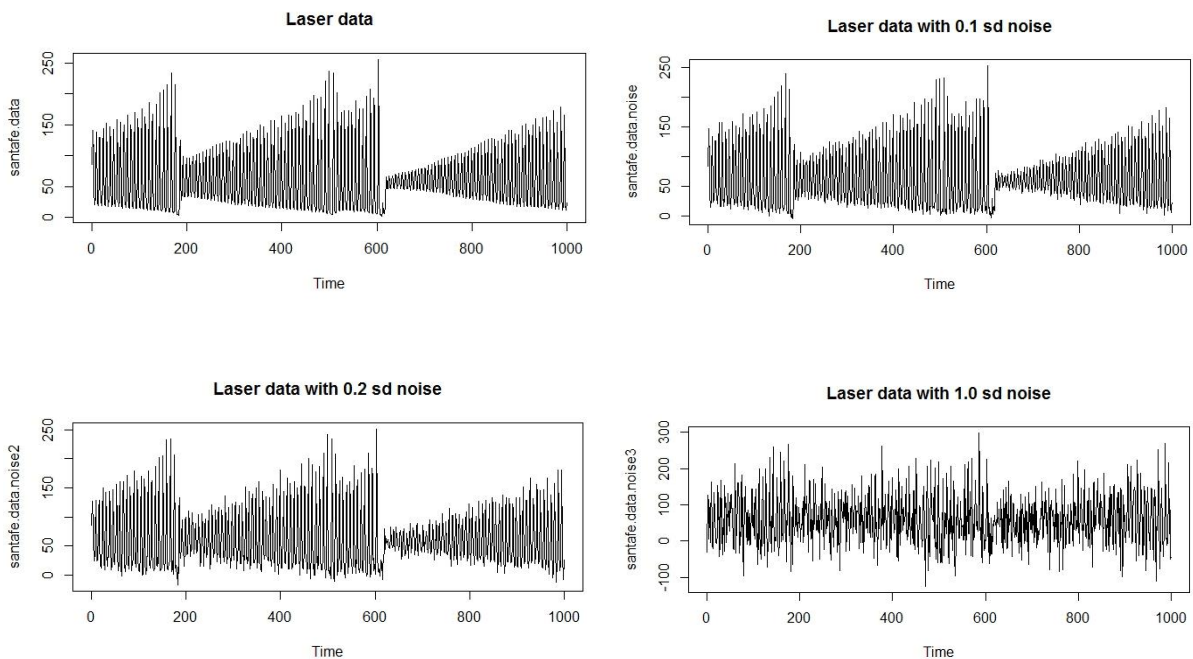

The original series exhibits heteroskedasticity and has varying regularity. Despite this, all four methods used were able to detect the increase in chaos with the addition of noise of increasing variance.

Figure 7 : Comparison for Santa Fe data

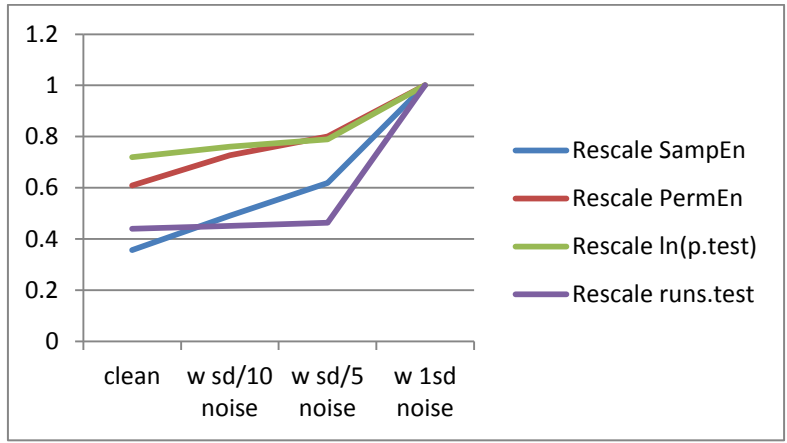

Table 3 : Scores for Santa Fe data

|  | clean | sd/10 noise | sd/5 noise | 1 sd noise |
|---|---|---|---|---|
| **SampEn(m=2)** | 0.7570 | 1.0441 | 1.3147 | 2.1233 |
| **PermEn(5)** | 0.5809 | 0.6933 | 0.7631 | 0.9529 |
| **Perm. test(5)** | 1562.7060 | 1045.5370 | 817.5509 | 198.3881 |
| **Perm. test(5) p-val** | 0.0000 | 0.0000 | 0.0000 | 0.0000 |
| **Runs test** | -15.3711 | -14.9967 | -14.6170 | -6.7707 |
| **Runs test p-val** | 0.0000 | 0.0000 | 0.0000 | 0.0000 |

The test statistics for all four measures show increasing chaos in the time series. However, the entropy statistic returned by SampEn and PermEn for the series with 1 standard deviation noise are so high that it would be difficult to distinguish between the noisy data and a purely random series. On the other hand, both the permutation test and the runs test have negligible p-values which would lead to rejection of the null hypothesis that the data originated from a random process.

MSE analysis was performed on the original clean series as well as the series with 0.2*standard deviation noise, with the results shown in Figure 8. As the original data is clearly periodic, it is again expected that the downsampling procedure will result in increasing scores as the scale factor increases, due to the reduction in correlations between successive points.

Figure 8

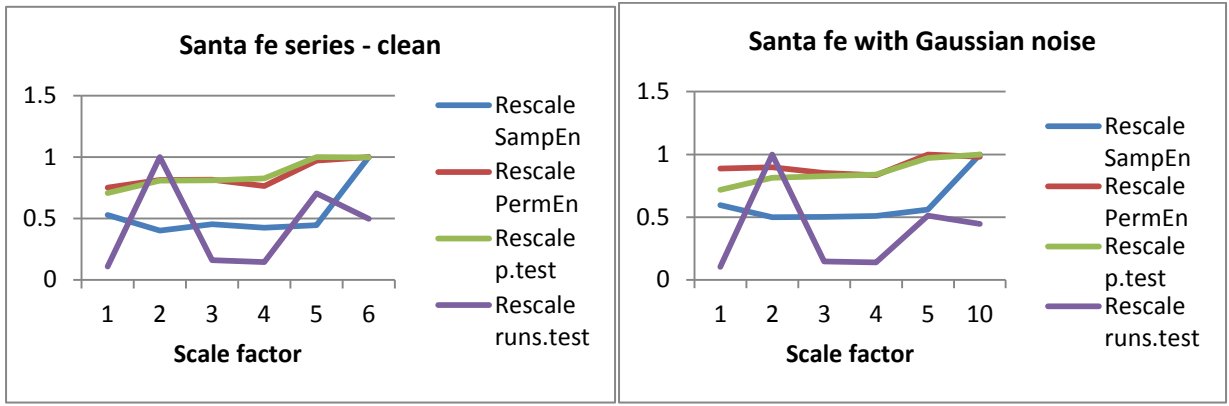

Table 4 : MSE analysis for Santa Fe data

| Santa fe - clean | | | | | | |
|---|---|---|---|---|---|---|
| Scale factor | 1 | 2 | 3 | 4 | 5 | 10 |
| SampEn(2) | 0.7570 | 0.5752 | 0.6507 | 0.6111 | 0.6406 | 1.4328 |
| PE(5) | 0.5809 | 0.6306 | 0.6316 | 0.5927 | 0.7528 | 0.7734 |
| Perm test(5) | 1562.7060 | 622.3253 | 610.2527 | 540.2703 | 181.9454 | 183.8897 |
| Perm test(5) p-val | 0.0000 | 0.0000 | 0.0000 | 0.0000 | 0.0002 | 0.0001 |
| Runs test | -15.3711 | 1.7011 | 10.5226 | 11.6606 | 2.4102 | 3.4173 |
| Runs test p-val | 0.0000 | 0.0889 | 0.0000 | 0.0000 | 0.0159 | 0.0006 |
| Santa fe with N(0,0.2*sd) noise | | | | | | |
| Scale factor | 1 | 2 | 3 | 4 | 5 | 10 |
| SampEn(2) | 1.3401 | 1.1230 | 1.1285 | 1.1460 | 1.2615 | 2.2513 |
| PE(5) | 0.7443 | 0.7534 | 0.7135 | 0.6995 | 0.8386 | 0.8236 |
| Perm test(5) | 939.9436 | 415.9501 | 381.2034 | 348.3164 | 157.9526 | 135.9184 |
| Perm test(5) p-val | 0.0000 | 0.0000 | 0.0000 | 0.0000 | 0.0098 | 0.1376 |
| Runs test | -14.8702 | 1.5221 | 10.3024 | 10.9002 | 2.9773 | 3.4173 |
| Runs test p-val | 0.0000 | 0.1280 | 0.0000 | 0.0000 | 0.0029 | 0.0006 |

Table 4 shows that the scores generally increase as the scale factor increases, indicating the detection of a loss in regularity caused by the downsampling process. While the runs test statistic again displays a large variation as the scale factor is varied, it has the advantage of returning a very low p-value even with a scale factor of 10, whereas the other 3 metrics return scores that are similar to those from a random series.

## 3.4 ARMA processes

The ARMA data set consists of 3 time series obtained by simulating several ARMA processes. The first series comes from an ARMA(2,2) process with AR coefficients of (0.9, -0.2) and MA coefficients of (-0.7, 0.1). The second series is an ARMA(1,1) process with an AR coefficient of 0.7 and MA coefficient of -0.2. The third and last series is an AR(1) process with coefficient 0.9. 1000 points were generated from each series, the plots which are shown below in Figure 8.

Figure 9 : Plots of varying ARMA processes

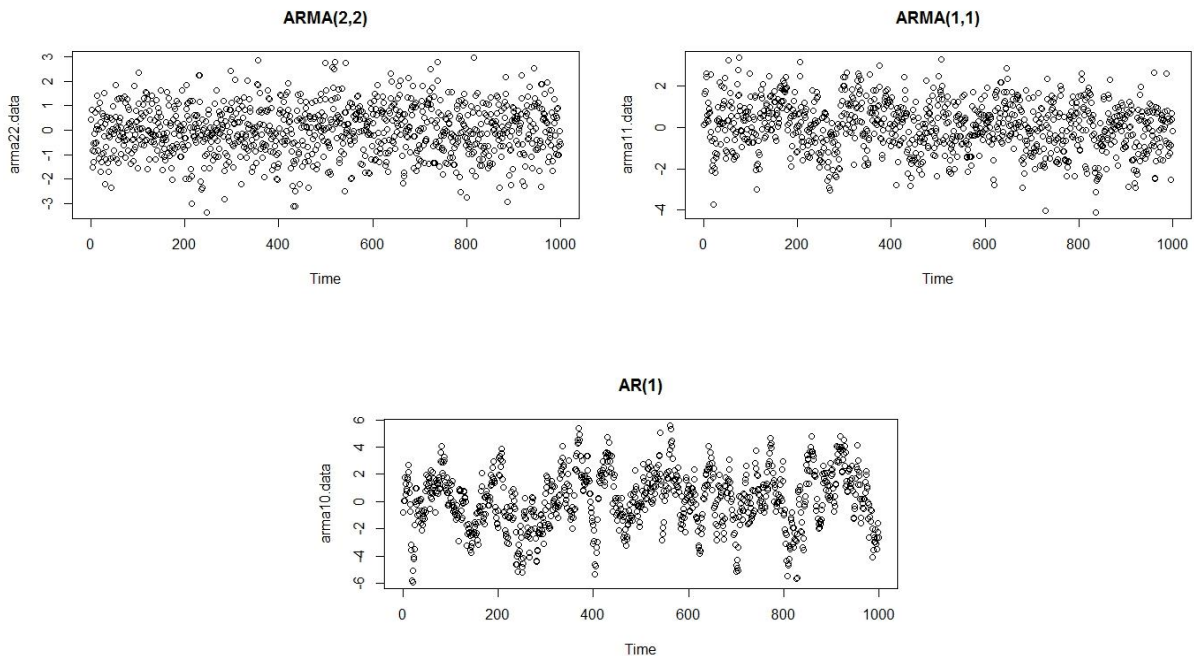

From inspection of the plots, increasing regularity can be seen when comparing the time series from the ARMA(2,2) process to the ARMA(1,1) process to the AR(1) process. Thus, the measures should show decreasing rescaled scores, as validated in Figure 10.

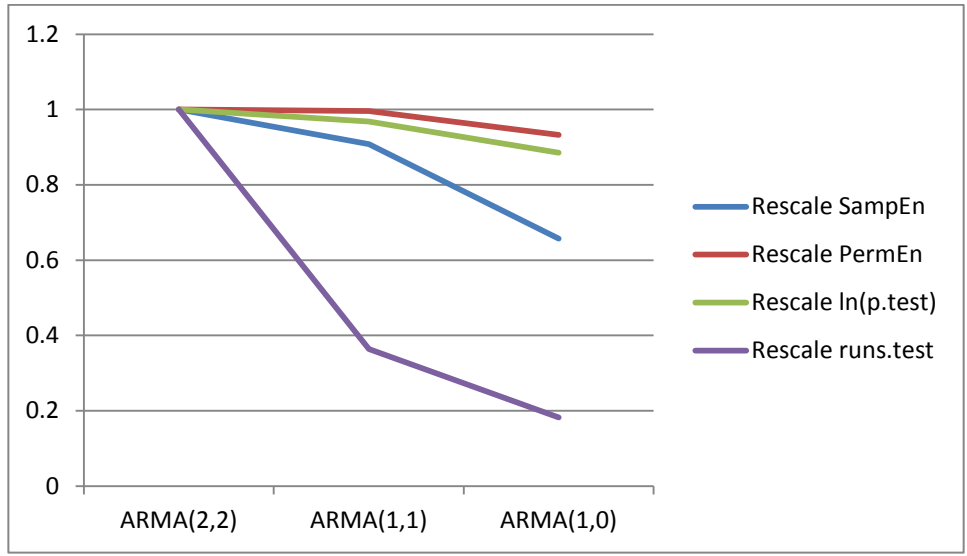

Figure 10 : Comparison for ARMA processes

Table 4: Score for ARMA data

|  | ARMA(2,2) | ARMA(1,1) | ARMA(1,0) |
|---|---|---|---|
| **SampEn(m=2)** | 2.2286 | 2.0238 | 1.4650 |
| **PermEn(5)** | 0.9833 | 0.9795 | 0.9173 |
| **Perm. test(5)** | 126.3924 | 147.9911 | 236.7858 |
| **Perm. test(5) p-val** | 0.3041 | 0.0369 | 0.0000 |
| **Runs test** | −3.9865 | −10.9470 | −21.8939 |
| **Runs test p-val** | 0.0001 | 0.0000 | 0.0000 |

The test statistics for all four measures indicate a decreasing measure of randomness or complexity. SampEn, PermEn and the permutation test are unable to distinguish between an ARMA(2,2) process and a purely random process, while the runs test is able to do so with a very small p-value. SampEn and PermEn still have problems with an ARMA(1,1) process, while the permutation test is able to reject the null hypothesis of a random process at a 0.05 significance level.

The MSE analysis of all 3 processes is shown in Figures 11a – 11c. Both the permutation test statistic and permutation entropy show minimal change as the scale factor increases, whereas there is significant variation in the sample entropy score and the runs test statistic.

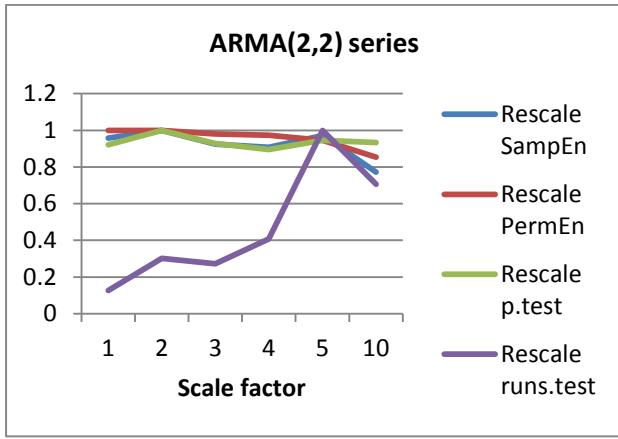

Figure 11a

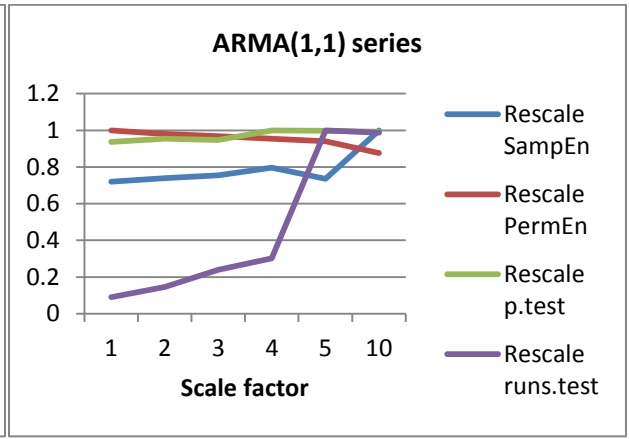

Figure 11b

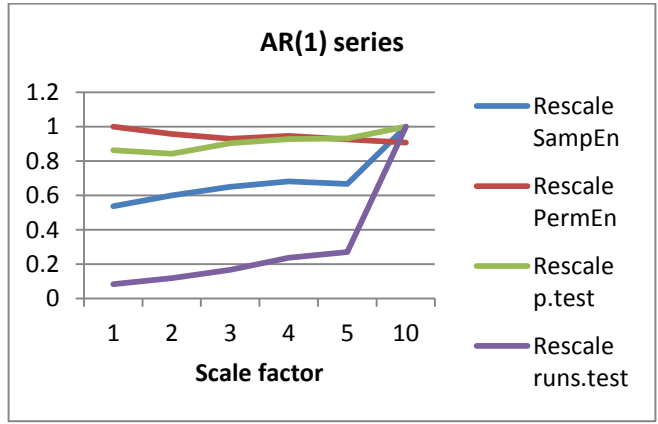

Figure 11c

Table 5 : MSE analysis for ARMA models

| ARMA(2,2) | | | | | | |
|---|---|---|---|---|---|---|
| Scale factor | 1 | 2 | 3 | 4 | 5 | 10 |
| SampEn(2) | 2.1533 | 2.2500 | 2.0808 | 2.0424 | 2.1864 | 1.7383 |
| PE(5) | 0.9846 | 0.9758 | 0.9559 | 0.9497 | 0.9235 | 0.8329 |
| Perm test(5) | 131.1921 | 89.5892 | 126.7042 | 151.5636 | 115.9652 | 123.9256 |
| Perm test(5) p-val | 0.2096 | 0.9797 | 0.2974 | 0.0235 | 0.5616 | 0.3601 |
| Runs test | -4.4927 | -1.8802 | -2.0824 | -1.3942 | 0.5671 | 0.8041 |
| Runs test p-val | 0.0000 | 0.0601 | 0.0373 | 0.1633 | 0.5706 | 0.4214 |
| ARMA(1,1) | | | | | | |

| Scale factor | 1 | 2 | 3 | 4 | 5 | 10 |
|---|---|---|---|---|---|---|
| SampEn(2) | 2.0238 | 2.0806 | 2.1228 | 2.2407 | 2.0680 | 2.8134 |
| PE(5) | 0.9795 | 0.9608 | 0.9486 | 0.9348 | 0.9217 | 0.8586 |
| Perm test(5) | 147.9911 | 135.1838 | 141.2470 | 108.3740 | 109.9670 | 111.9328 |
| Perm test(5) p-val | 0.0369 | 0.1474 | 0.0802 | 0.7476 | 0.7114 | 0.6642 |
| Runs test | -10.9470 | -6.8045 | -4.1648 | -3.2954 | -0.9924 | -1.0051 |
| Runs test p-val | 0.0000 | 0.0000 | 0.0000 | 0.0010 | 0.3210 | 0.3149 |
| AR(1) | | | | | | |
| Scale factor | 1 | 2 | 3 | 4 | 5 | 10 |
| SampEn(2) | 1.4650 | 1.6371 | 1.7735 | 1.8615 | 1.8197 | 2.7300 |
| PE(5) | 0.9173 | 0.8782 | 0.8527 | 0.8691 | 0.8503 | 0.8322 |
| Perm test(5) | 236.7858 | 269.5677 | 184.8755 | 161.1613 | 157.9526 | 111.9328 |
| Perm test(5) p-val | 0.0000 | 0.0000 | 0.0001 | 0.0061 | 0.0098 | 0.6642 |
| Runs test | -21.8939 | -15.2205 | -10.8504 | -7.6048 | -6.6636 | -1.8092 |
| Runs test p-val | 0.0000 | 0.0000 | 0.0000 | 0.0000 | 0.0000 | 0.0704 |

Examination of table 5 above reveals that similar to the case with random data, the permutation entropy score decreases almost monotonically as scale factor increases. Sample entropy and the permutation test statistic do not display any obvious trend with the change in scale factor.

### 3.5    Congestive Heart Failure – Normal Sinus Rhythm data

Entropy measures are commonly used in the analysis of physiologic time series. For the final evaluation of entropy measures against tests for randomness, two separate inter-beat (RR) interval time series were evaluated[1]. The first series is comprised of measurements from 44 patients with congestive heart failure (CHF)[10], while the second series is comprised of measurements from 54 patients with normal sinus rhythm (NSR). The series length varies from 75,546 data points to 147,880 data points for the CHF group, and from 76,927 data points to 136,528 data points for the NSR group.

SampEn, PermEn, the permutation test and the runs test were applied to each of the time series in both groups. Similar to the previous experiments, for SampEn the parameters m=2 and r=0.2 was chosen, while for PermEn and the permutation test the tuple size was set to 5. Figures 12a-d show the results of the various measures on CHF patients (red box plot) and NSR patients (blue box plot).

---

[1] Provided by Douglas E. Lake, UVA Department of Medicine

Figure 12a : SampEn          Figure 12b : PermEn

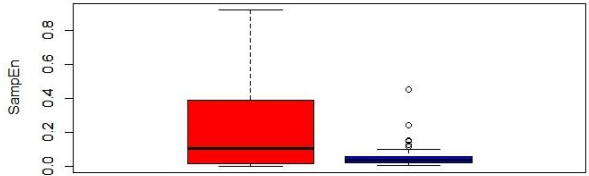          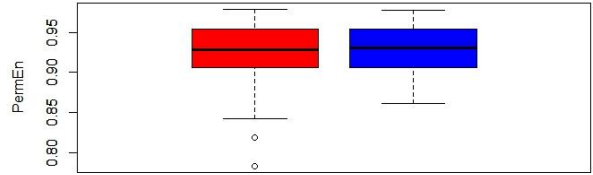

Figure 12c : Permutation test          Figure 12d : Runs test

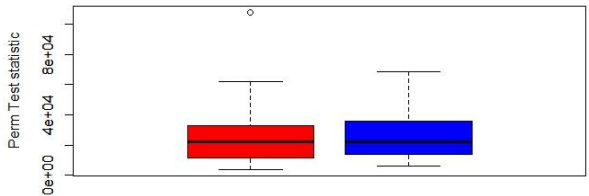          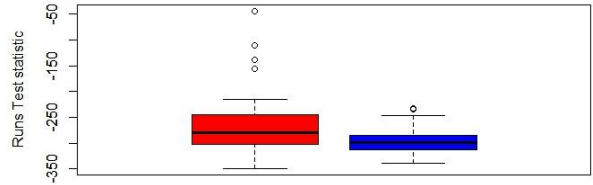

It is readily apparent that SampEn scores are markedly different between the two groups, with a lower mean score and much lower variance. This suggests that the NSR patients have more regular inter-beat interval times. The scores for PermEn and the permutation test are not markedly different between the two groups, again highlighting the similarity between these two measures. Lastly, the runs test statistic provides similar results as the SampEn score, except to a lesser degree. The absolute value of the runs test statistic for NSR patients is larger, again suggesting that NSR patients have inter-beat interval times that are more regular. The variance of the runs test statistic is also lower, similar to what was found for SampEn.

A two-sample t-test of the scores from both groups quantifies the difference shown in the box plots. For both SampEn and the runs test, we are able to reject the null hypothesis that the two samples have the same mean at a 0.05 significance level, whereas for PermEn and the permutation test we are unable to reject the null hypothesis at a 0.05 significance level.

## 4. Discussion and future work

Entropy based methods such as sample entropy and permutation entropy are able to quantify the degree of regularity present in a series and utilize this measure as a means of comparing the complexity of different time series. In a similar fashion, established tests for randomness such as the permutation test and the runs test examine a series for the presence of underlying structure in order to determine the likelihood of the series originating from a random distribution.

Experimental analysis shows that the test statistics of the permutation test and the runs test vary in a fashion that is highly correlated to SampEn and PermEn scores. Thus, these tests may be able to provide similar information regarding the complexity of time series when comparing multiple data sets. Furthermore, such statistical tests also have the advantage of having well-known statistical distributions which can provide probabilistic information on the likelihood of the originating process being random. In comparison, a measure of entropy from SampEn and PermEn by itself may not provide enough information to make the aforementioned distinction, even with the application of the MSE framework. In some cases, the p-values from the permutation test and the runs test can provide additional information that is not detectable by changes in SampEn and PermEn.

Further study should be carried out on the potential applications of various other tests for randomness in conjunction with entropy-based measures to gain further insight into the complexity of time series, which may provide additional predictive power.